\pdfoutput=1
\documentclass[11pt]{article}
\usepackage{pdfpages}
\usepackage{amssymb}
\usepackage{amsmath}
\usepackage{amsfonts}
\usepackage{enumerate}
\usepackage{amsthm}
\usepackage{moreverb}
\usepackage{subcaption}
\usepackage{float}
\usepackage{geometry}
\usepackage{natbib}

\newcommand{\asim}{ \stackrel{\mbox{\small $a$}}{\sim}}

\begin{document}

\title{{\Large  Confidence Intervals for Quantiles from Histograms and Other Grouped Data}
\author{Dilanka S. Dedduwakumara, Luke A. Prendergast \\
Department of Mathematics and Statistics\\ 
 La Trobe University
 }
 } 
\date{} 
\maketitle

\begin{center}
\textbf{Abstract}
\end{center}
Interval estimation of quantiles has been treated by many in the literature. However, to the best of our knowledge there has been no consideration for interval estimation when the data are available in grouped format. Motivated by this, we introduce several methods to obtain confidence intervals for quantiles when only grouped data is available. Our preferred method for interval estimation is to approximate the underlying density using the Generalized Lambda Distribution (GLD) to both estimate the quantiles and variance of the quantile estimators. We compare the GLD method with some other methods that we also introduce which are based on a frequency approximation approach and a linear interpolation approximation of the density. Our methods are strongly supported by simulations showing that excellent coverage can be achieved for a wide number of distributions.  These distributions include highly-skewed distributions such as the log-normal, Dagum and Singh-Maddala distributions. We also apply our methods to real data and show that inference can be carried out on published outcomes that have been summarized only by a histogram.  Our methods are therefore useful for a broad range of applications.  We have also created a web application that can be used to conveniently calculate the estimators.

\vspace*{.3in}

\newpage

\section{Introduction} \label{Introduction}

Grouped data refers to data that has been aggregated into groups.  When the data are continuous, a common approach is to segment the range of the data into a finite number of non-overlapping intervals (or \textit{bins}). Frequencies indicating the number of individuals in each interval are then tallied and the intervals and frequencies together are commonly referred to as a frequency distribution.  It is this scenario of grouped data that is our focus and, in particular, we are interested in cases where the raw data are unavailable.   

There are many reasons why data may have been made available as grouped data and as to why the raw data may not be available. For example, government agencies may only release grouped data to preserve confidentiality of individuals or a researcher may only have at their disposal a histogram from previous published findings.

In situations such as the above, obtaining reliable interval estimates for quantiles may be extremely useful. In this paper we demonstrate several such methods to obtain confidence intervals from grouped data. Our methods estimate the underlying empirical distribution using the grouped data and together with asymptotic normality results we obtain our intervals. What is quite surprising is just how reliable these intervals can be even when the raw data sample size or number of bins is not large.  These methods can be easily implemented in statistical packages like R, SAS etc. and some of the methods can even be implemented in Excel.       

\subsection{Motivating Example} \label{Motivating Example}

Here we consider two histograms depicting Glucose effectiveness for young and healthy Caucasian men ($n=186$) and women ($n=194$).  These histograms were published by \cite{clausen1996insulin} and are provided in their original form below in Figure \ref{fig1}. One hypothesis that is of interest is whether there is a difference between men and women with regards to glucose effectiveness.  

\begin{figure}[H]
\begin{subfigure}{0.5\textwidth}
  \centering
  \includegraphics[width=.8\linewidth]{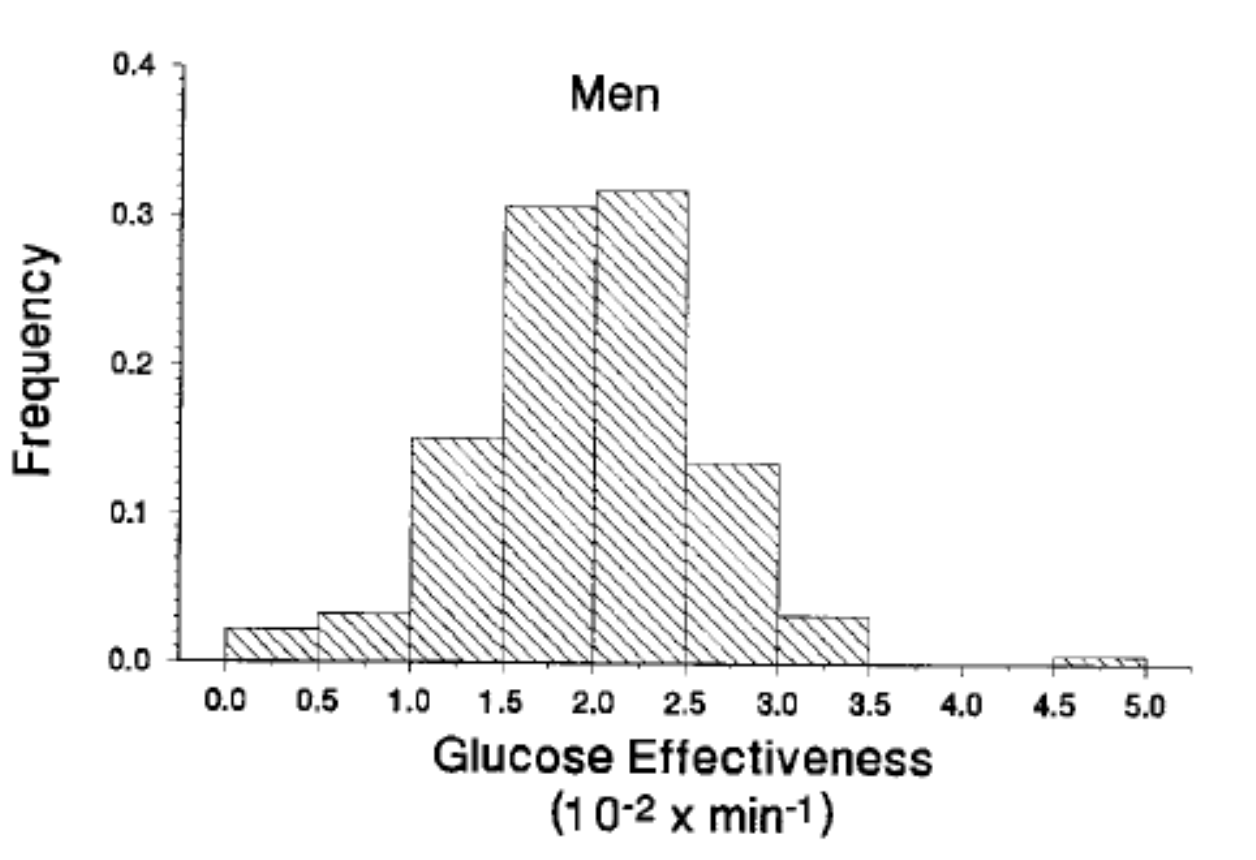} 
\end{subfigure}%
\begin{subfigure}{.5\textwidth}
  \centering
  \includegraphics[width=.8\linewidth]{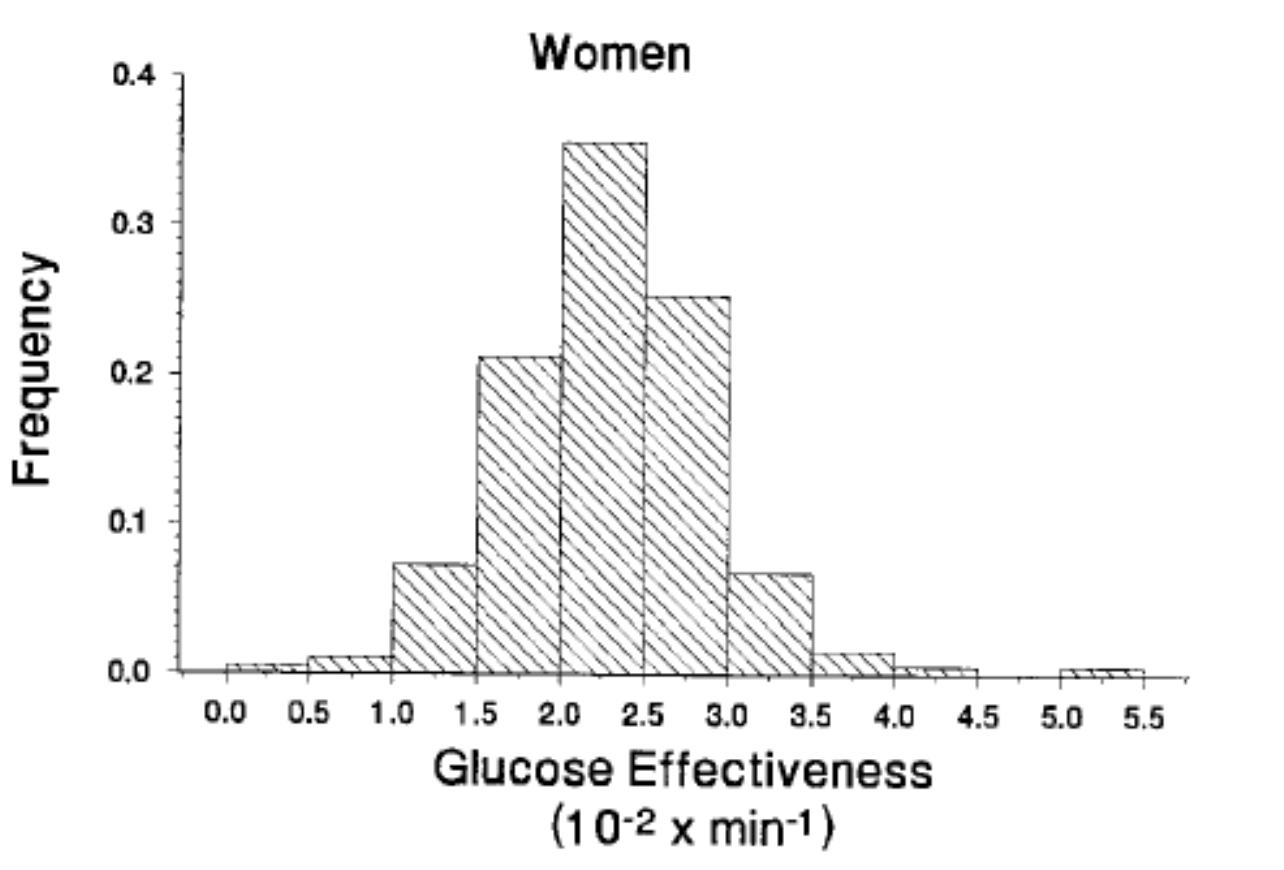}  
\end{subfigure}
\caption{Histograms of measures of glucose effectiveness conditional on gender. Reprinted from "Insulin sensitivity index, acute insulin response, and glucose effectiveness in a population-based sample of 380 young healthy Caucasians. Analysis of the impact of gender, body fat, physical fitness, and life-style factors," by J.O.Clausen, K.B.Johnsen, H.Ibsen, R.N.Bergman, P.Hougaard, K.Winther and O.Pedersen, Journal of Clinical Investigation, 98, p. 1195. Copyright 1996 by American Society for Clinical Investigation, Inc. Reprinted with permission.}

\label{fig1}
\end{figure}

The question then is, \textit{without the raw data can we conduct a statistical inference to investigate this hypothesis based only on these histograms?}  The answer is \textit{yes}.   We will show later that our methods can be used to obtain reliable confidence intervals for quantiles and for the difference of quantiles.  

\subsection{Asymptotic Theory for Quantile Estimators} \label{Asymptotic Theory for Quantiles}

The asymptotic theory for sample quantile estimators has been treated and discussed over the past several decades. Fundamental results for asymptotic theory for quantiles can be found in many places;   \cite{serfling2009approximation}, \cite{ferguson1996course}, \cite{gilchrist2000statistical} and \cite{dasgupta2008asymptotic} are a few examples. Below we provide the asymptotic variance for quantile estimators that can be found in the aforementioned works. 

Let $f$, $F$ and $Q$ denote the density, distribution and quantile functions respectively.  For $p \in [0,1]$, let  
$x_p=Q(p)=F^{-1}(p)$ denote the $p$-th quantile or, equivalently, the $p\times 100$th percentile.  Given a random sample of data denoted $X_1,\ldots,X_n$ and letting $F_n$ denote the empirical distribution for this data, then an estimator of $x_p$ is $\widehat{x}_p=F_n^{-1}(p)$.  Then, see e.g. Chapter 7 of \cite{dasgupta2008asymptotic}, for suitably large $n$,  
\begin{equation}
\sqrt{n}\widehat{x}_p \asim N \left(x_p,\frac{p(1-p)}{f^2(x_p)} \right)
\label{eq1}
\end{equation}
where $\asim$ denotes `approximately distributed as'.

The result in \eqref{eq1} allows us to construct an approximate $100(1-\alpha)\%$ confidence interval for $x_p$ as follows,
\begin{equation}
\widehat{x}_p \pm z_{1-\alpha/2}\sqrt{p(1-p)/\left[n\widehat{f}^2(\widehat{x}_p)\right]} 
\label{eq2}
\end{equation}
where $z_{1-\alpha/2}=\Phi^{-1}(1-\alpha/2)$ denotes the $1-\alpha/2$ quantile of the $N(0,1)$ distribution and $\widehat{f}$ denotes an estimate for the density $f$.  

The interval in \eqref{eq2} can be easily extended to compare independent quantile estimators as follows.  Let $\widehat{x}_p$ and $\widehat{y}_p$ be the $p$ quantile estimates based on two independent random samples of size $n$ and $m$  respectively.  For $\widehat{f}$ and $\widehat{g}$ denoting the density estimates, a large sample confidence interval for the difference of the quantiles is,
\textbf{ \begin{equation}
(\widehat{x}_p - \widehat{y}_p) \pm z_{1-\alpha/2}\ \sqrt[]{\frac{p(1-p)}{n\widehat{f}^2(\widehat{x}_p)} + \frac{p(1-p)}{m\widehat{g}^2(\widehat{y}_p)}}. 
\label{eq3}
\end{equation}} 

When the raw data are available, there are many methods that may be employed to estimate the unknown density and a nice overview of several methods is given by \cite{sheather04}.  In the next section we restrict our attention to density estimation when grouped data are available.

\section{Methods} \label{methods}

In this section we present our methods which derive the estimates $\widehat{x}_p$ and $\widehat{f}(\widehat{x}_p)$ from available bin points and frequencies.  In doing so we subsequently construct the confidence intervals for quantiles of the form in \eqref{eq2}. 

\subsection{GLD Estimation method} \label{GLD Estimation method}

The GLD distribution is often encountered in fields such as finance due to its ability to approximate a large number of well-known distributions. The four parameter GLD distribution consists of a location parameter $\lambda$, an inverse scale parameter $\eta$ and two non linear shape parameters $\alpha$ and $\beta$. The FKML parameterization \cite{freimer1988study} of the GLD distribution, which  will be our focus here, is defined in terms of its quantile function given as
\begin{equation*}
Q(p)=\lambda + \frac{1}{\eta} \bigg[\frac{(p^\alpha-1)}{\alpha}-\frac{\{(1-p)^\beta-1\}}{\beta}\bigg].
\label{eq4}
\end{equation*}

Due to its flexibility, the GLD distribution has been used recently to obtain confidence intervals for quantiles \cite[e.g.]{su2009confidence,prendergast2016exploiting}, although not with respect to grouped data.  However, methods are available to estimate the parameters of the GLD distribution for grouped data using the frequencies and the break points of the bins. We estimate the parameters using the so-called Percentile Matching method discussed in \cite{karian1999fitting} and \cite{tarsitano2005estimation} which equates a selected set of approximated empirical percentiles with their corresponding theoretical counterparts. By equating those percentiles, the following system of nonlinear equations can be specified:
\begin{equation*}
\widehat{x}_{p}=\widehat{\lambda} + \frac{1}{\widehat{\eta}} \bigg[\frac{(p^{\widehat{\alpha}}-1)}{\widehat{\alpha}}-\frac{\{(1-p)^{\widehat{\beta}}-1\}}{\widehat{\beta}}\bigg] \quad for \quad p \in {[10\%,25\%,50\%,75\%,90\%]}
\label{eq5}
\end{equation*}
where $\widehat{x}_p$ denotes the approximated empirical percentiles from the binned data. \\

Using the optimization method L-BFGS-B \cite{byrd1995limited}, the system of nonlinear equations can be optimized to obtain parameter estimates for the GLD distribution. This method is readily available in  the "bda" package \cite{wang2015bda} in the R statistical software \cite{R} which enables the GLD density estimation based on grouped (or pre-binned) data. Estimated parameters can then be used to determine the estimated GLD quantile $\widehat{x}_p$ and the estimated GLD density $\widehat{f}(\widehat{x}_p)$, which can then be used to construct confidence intervals as in (\ref{eq2}) for $x_p$. 

\subsection{Histogram method} \label{histogram method}

A histogram is the simplest way of estimating (or approximating) a probability density function once the area has been scaled so that it is equal to one.  Density estimation using histograms has been previously looked at by \cite{tille2012histogram} in estimating the
Lorenz curve and Gini index commonly used in studies of inequality. 

A rough estimate of the quantile function can be obtained from the information available within a histogram by applying a simple interpolation as below.    
\begin{equation*}
\widehat{x}_p=l+\frac{hN(p-p_{close})}{n_p} 
\label{eqn:eq6}
\end{equation*}
where
\begin{enumerate}
\item[] $l$= The lower class boundary of the class containing $\widehat{x}_p$.

\item[] $h$= The width of the class containing $\widehat{x}_p$.

\item[] $N$= The total frequency.

\item[] $p_{close}$= Cumulative probability of the class immediately preceding to the class containing $\widehat{x}_p$.

\item[] $n_p$= Class frequency of the class containing the $\widehat{x}_p$.

\end{enumerate}

Then the corresponding density estimate for $x_p$ is found by,

\begin{equation*}
\widehat{f}(\widehat{x}_p)= \frac{n_p}{hN}
\label{eqn:eq7}
\end{equation*}

Substitution of these estimated values in (\ref{eq2}) can then be used to obtain confidence intervals for $x_p$.

\subsection{Linear interpolation method } \label{linear interpolation method}

\cite{lyon2016advantages} introduced a linear interpolation method to estimate the density for grouped data when the additional information of group means are available. This method assumes a linear density for each bin except for the final bin where an exponential tail is fitted and therefore allowed to be unbounded. Closed form solutions are obtained for the density function.

Assume that the sample values are grouped into \textit{J} intervals defined by the bounds $[a_{j-1},a_j), j=1,\ldots,J$. These values are bounded below with $a_0>-\infty$ and unbounded in the final interval with $a_J=\infty$. The relative frequency for the $j$th group is denoted by $\widehat{f_j}$ and the cumulative relative frequency up to \textit{j}th interval is denoted by $\widehat{F_j}$ . The mean of the $j$th group is $\bar{x_j}$ and the midpoint of the group by $x^c_j$. When $p\in [0,1)$, the density function estimate is defined by the linear equation
\begin{equation*}
h_j(x)=\alpha_j+\beta_j x, \qquad x\in [a_{j-1},a_j)
\label{eq8}
\end{equation*}
where $\alpha_j$, $\beta_j$ are the linear density coefficients. \cite{lyon2016advantages} estimate these coefficients by  
\begin{equation*}
\notag
\widehat{\beta_j}=\widehat{f_j}\frac{12(\bar{x_j}-x^c_j)}{(a_j-a_{j-1})^3}, \quad 
\widehat{\alpha_j}=\frac{\widehat{f_j}}{a_j-a_{j-1}}-\widehat{\beta_j}x^c_j.
\end{equation*}

The density estimate for the exponential tail assumed for the last (unbounded) interval is
\begin{equation*}
h_J(x)=\frac{\eta}{\lambda}\exp\left\{\frac{(x-a_{J-1})}{\lambda}\right\}
\label{eq9}
\end{equation*}
where the parameters are estimated as,
$\widehat{\eta}=\widehat{f_J}$ and $\widehat{\lambda}=\bar{x_J}-a_{J-1}$.

Using the above estimated coefficients, the closed form solution of the quantile function for the bounded interval in $[a_{j-1},a_j)$ can be derived as,
\begin{equation*}
\widehat{x}_p=\frac{-\widehat{\alpha}_j + \sqrt[]{2\widehat{\beta}_jp+\widehat{C}_j}}{\widehat{\beta}_j}
\label{eq10}
\end{equation*}
where, 
\begin{equation*}
\notag
\widehat{C}_j=[\widehat{\alpha}_j^2-2\widehat{\beta}_j\widehat{F}_{j-1}+2\widehat{\beta}_j\widehat{\alpha}_ja_{j-1}+\widehat{\beta}_j^2(a_{j-1})^2].
\end{equation*}

The fitted exponential tail will yield the following quantile function for the unbounded interval
\begin{equation*}
\widehat{x}_p=a_{J-1}-\widehat{\lambda}\ln\left(1-\frac{p-\widehat{F}_{J-1}}{\widehat{\eta}}\right).
\label{eq11}
\end{equation*}

The asymptotic confidence interval for $x_p$ can be constructed by substituting the estimated quantile function and the interpolated density estimate into \eqref{eq2}. 

\subsection{Frequency polygon method} \label{frequency polygon method}

The frequency polygon method is an extension of the histogram method in which the midpoint of the bin values are connected by line segments. This adds a sense of continuity and smoothness for the histogram.  Motivated by the frequency polygon methods considered in \cite{scott1985frequency} and the linear interpolation method above, we also consider allowing for linear change between bin midpoints but where the estimated line segments connect.  This method does not require bin means, but in what follows we do consider equal bin widths. 

Let $J$ again be the number of intervals of the histogram where the midpoints of the bins are denoted as $x_1^c, \ldots, x^c_J$.  Further, let the histogram density estimates be $g_j$ where $j=1,\ldots,J$, and the  bin width of the histogram, assumed equal for all bins, to be $h$.  Since the frequency polygon has different breakpoints compared to the histogram, they are defined by the bounds $[a_{k-1},a_k), k=1,\ldots,K$ where $K=J+1$. These breakpoints are related to the histogram bins as, $\{a_0,\ldots,a_K\}$ = $\{ x^c_1-h,x^c_1,\ldots,x^c_{J},x^c_{J}+h\}$. Further, the corresponding density estimates of each break point of frequency polygon is defined by $\{f_0,\ldots,f_K\}$ = \{ $0,g_1,...,g_J,0$\} . By using the same methodology discussed above in linear interpolation method, density function from the frequency polygon method can be defined as follows,

\begin{equation*}
\widehat{h}_k(x)=\widehat{\alpha}_k+\widehat{\beta}_k x, \qquad x\in [a_{k-1},a_k)
\label{eq12}
\end{equation*}
where
\begin{equation}
\notag
\widehat{\beta}_k=\frac{f_k-f_{k-1}}{a_k-a_{k-1}} \;\;\text{and}\;\; 
\widehat{\alpha}_k=f_{k-1}-\widehat{\beta}_k a_{k-1}.
\end{equation}

Closed form solutions for the quantile function of Frequency polygon method can be obtained in the same way discussed in \cite{lyon2016advantages} for the linear interpolation method. 

\subsection{Estimation} \label{Estimation}
\begin{figure}[h]
\centering
\includegraphics[height=9cm,width=\textwidth]{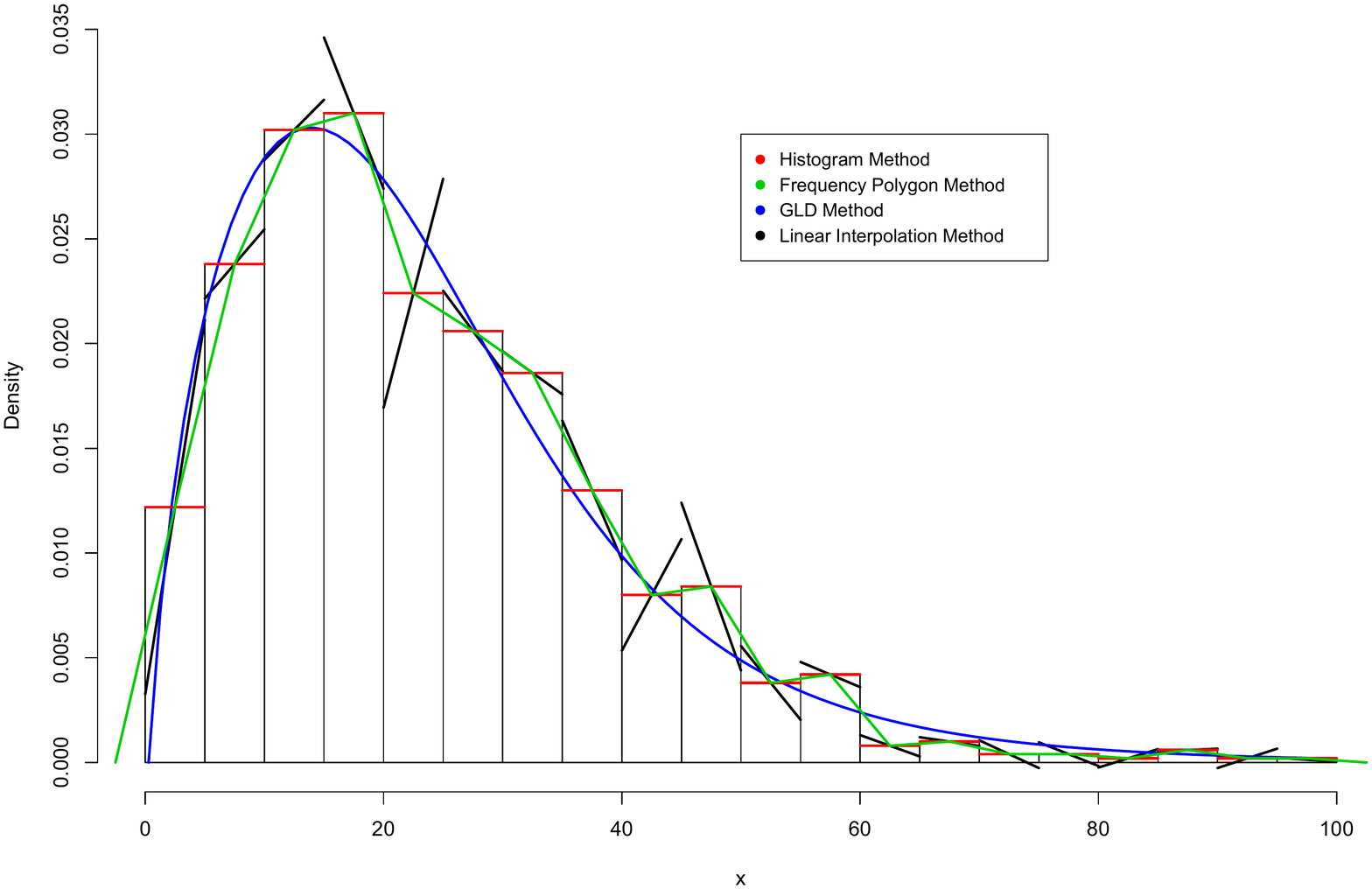}
\caption{ Estimation of the density from histogram of 1000 observations from Singh-Maddala distribution.}
\label{fig2}
\end{figure}

\begin{figure}[h]
\centering
\includegraphics[height=11cm,width=\textwidth]{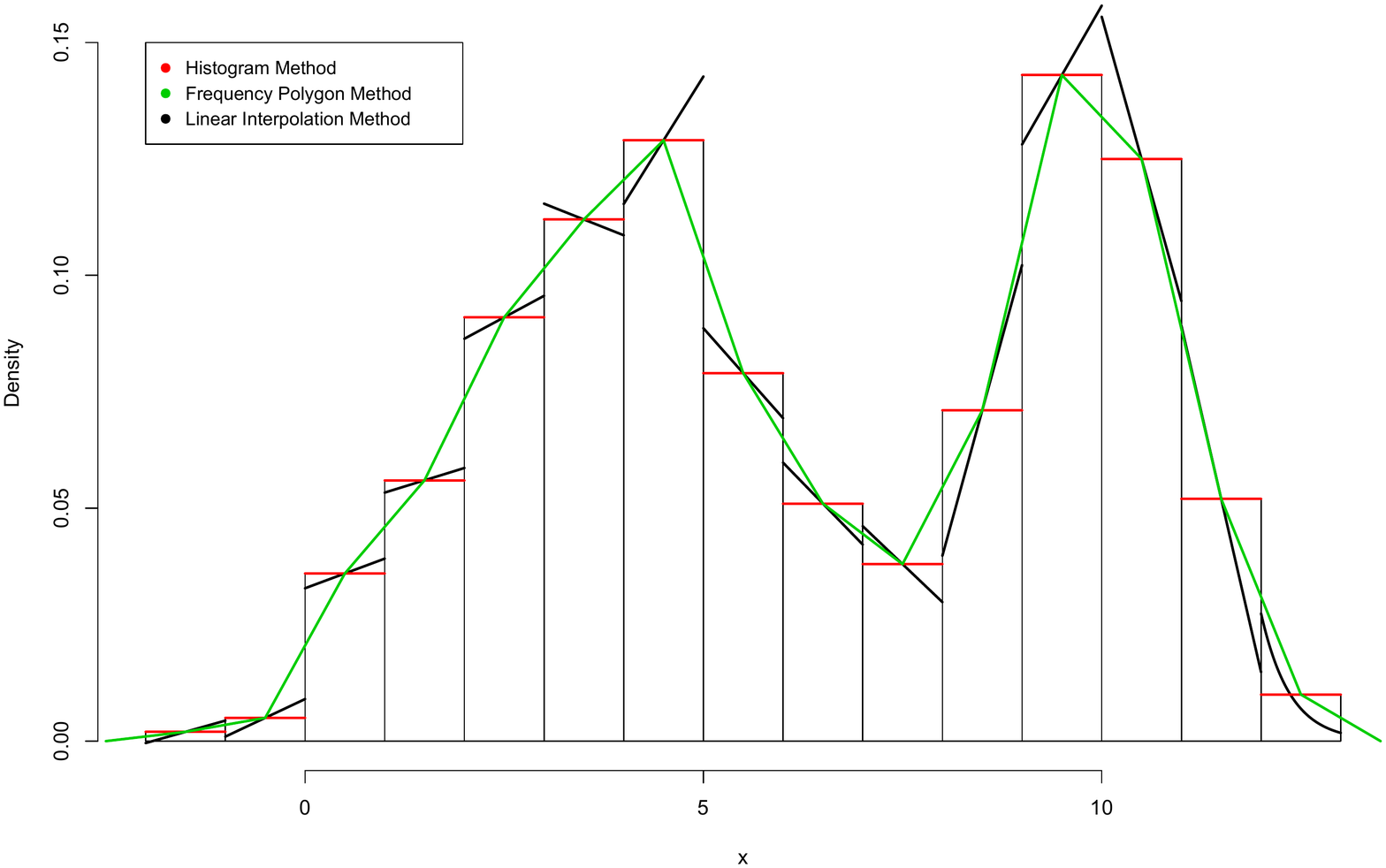}
\caption{Estimation of the density from a histogram of 1000 observations sampled from a normal mixture distribution.}
\label{fig3}
\end{figure}

In Figures \ref{fig2} and \ref{fig3} we provide estimates of the density using the above methods from a unimodal and bimodal histogram respectively obtained from simulated data sets. The group means which are needed for the linear interpolation method are calculated using the full simulated data set before binning. 

For the unimodal histogram shown in Figure \ref{fig2}, 1000 observations were randomly sampled from the Singh-Maddala Distribution. We chose this distribution since it has useful applications such as the modeling of skewed income data. The overlaid estimated GLD density and the linear interpolation density capture the shape of the histogram data well compared to the other methods . The uniformity within a bin from the histogram method means that it may not provide good approximations in domains for which the gradient of the density is steep. The frequency polygon method may improve on this by allowing for the linear change within bins. 

For the bimodal scenario, we again generated 1000 observations but this time from a normal mixture distribution.  Here, the GLD method cannot be used since it is a unimodal distribution and therefore should only be used to estimate unimodal densities.  However, all other methods provide a reasonable fit for this bimodal distribution as can be seen in Figure \ref{fig3}.

\section{Simulations and Examples} \label{Simulations and Examples} 

\subsection{Simulations} \label{Simulations}

In this section we apply our methods to simulated grouped data from several distributions and assess their effectiveness by looking at coverage probabilities and interval widths. While it occurs rarely, bins which are of zero counts are merged with others to make sure that density estimates are non zero.  The $(1-\alpha/2)\times 100\%$ confidence interval for $x_p$ is estimated as $\hat{x}_p \pm z_{1-\alpha/2}\sqrt[]{p(1-p)/\big[ \hat{f}^2(x_p)n \big] }$ where $\hat{x_p}$ and $\hat{f}(\widehat{x}_p)$ are computed using the methods from the previous section. The variation of the coverage for each $x_p$ for $p \in \{0.1,0.25,0.5,0.75,0.9\}$ are explored for several choices of sample size and some commonly encountered distributions.

The log-normal distribution with $\mu=0$ and $\sigma=0.25$ and Singh-Maddala distribution with parameter values, $a=1.6971$, $b=87.6981$ and $q=8.3679$  were  considered. Parameters for the Singh-Maddala were from fitted US family incomes in 1980 reported in \cite{mcdonald1984some}. For the Dagum distribution, the parameter choices of $a=4.273$ $b=14.28$ and $p=0.36$ were used in \cite{kleiber2008guide} which were also estimated from fitted US family incomes, this time from 1969. We also considered the standard normal distribution as a symmetric distribution.

\begin{table}[H]
  \centering
  
    \caption{Empirical coverage probabilities and average widths of interval estimates of $x_p$ from binned data at nominal level 95\% for moderate sample sizes,each based on 1000 replications.}
\label{table:tab1}
\hspace*{-1cm}
\begin{tabular}{ccccccc}
\hline
\textit{n} & \textit{F} & \textit{p} & Histogram & Frequency Polygon & Linear Interpolation & GLD\\
\hline
&  & 0.10 & 0.907  (0.209) & 0.962  (0.199) & 0.905  (0.184) & 0.895  (0.197)\\
&  & 0.25 & 0.965  (0.165) & 0.969  (0.179) & 0.894  (0.157) & 0.948  (0.173)\\
& Lognormal & 0.50 & 0.937  (0.177) & 0.962  (0.186) & 0.930  (0.176) & 0.965  (0.180)\\
&  & 0.75 & 0.938  (0.223) & 0.947  (0.231) & 0.916  (0.216) & 0.951  (0.228)\\
&  & 0.90 & 0.905  (0.312) & 0.921  (0.336) & 0.863  (0.414) & 0.943  (0.378)\\
\cline{2-7}
 &  & 0.10 & 0.907  (4.233) & 0.953  (4.15) & 0.937  (3.483) & 0.898  (4.382)\\
 &  & 0.25 & 0.977  (3.777) & 0.982  (4.145) & 0.942  (3.644) & 0.952  (3.959)\\
 & Dagum & 0.50 & 0.945  (4.189) & 0.960  (4.479) & 0.937  (3.873) & 0.954  (4.304)\\
 &  & 0.75 & 0.949  (5.119) & 0.949  (5.551) & 0.903  (5.011) & 0.948  (5.640)\\
 &  & 0.90 & 0.920  (9.772) & 0.924  (9.213) & 0.871  (10.43) & 0.908  (9.643)\\
\cline{2-7}
50 &  & 0.10 & 0.898  (9.028) & 0.952  (8.623) & 0.943  (7.12) & 0.886  (8.637)\\
 &  & 0.25 & 0.969  (8.579) & 0.974  (9.119) & 0.899  (8.064) & 0.949  (8.766)\\
 & Singh-Maddala & 0.50 & 0.946  (10.32) & 0.963  (10.67) & 0.932  (11.01) & 0.971  (10.39)\\
 &  & 0.75 & 0.908  (14.34) & 0.949  (14.59) & 0.888  (13.80) & 0.961  (14.54)\\
 &  & 0.90 & 0.814  (22.69) & 0.912  (22.56) & 0.801  (31.82) & 0.916  (24.89)\\
\cline{2-7}
 &  & 0.10 & 0.889  (1.071) & 0.934  (0.996) & 0.827  (1.478) & 0.904  (1.021)\\
 &  & 0.25 & 0.925  (0.769) & 0.958  (0.792) & 0.888  (0.730) & 0.959  (0.773)\\
 & Normal & 0.50 & 0.937  (0.722) & 0.952  (0.739) & 0.929  (0.762) & 0.962  (0.723)\\
 &  & 0.75 & 0.924  (0.767) & 0.957  (0.790) & 0.871  (0.725) & 0.960  (0.779)\\
 &  & 0.90 & 0.798  (0.950) & 0.935  (0.989) & 0.811  (1.066) & 0.868  (0.995)\\
\cline{1-7}
 &  & 0.10 & 0.885  (0.145) & 0.930  (0.142) & 0.936  (0.124) & 0.924  (0.145)\\
 &  & 0.25 & 0.975  (0.113) & 0.973  (0.126) & 0.935  (0.113) & 0.950  (0.123)\\
 & Lognormal & 0.50 & 0.932  (0.127) & 0.963  (0.133) & 0.940 (0.121) & 0.955  (0.126)\\
 &  & 0.75 & 0.968  (0.151) & 0.972  (0.164) & 0.925  (0.162) & 0.967  (0.160)\\
 &  & 0.90 & 0.934  (0.216) & 0.948  (0.235) & 0.902  (0.396) & 0.965  (0.251)\\
\cline{2-7}
 &  & 0.10 & 0.819  (2.998) & 0.848  (3.005) & 0.971  (2.541) & 0.857  (3.126)\\
 &  & 0.25 & 0.962  (2.654) & 0.966  (2.967) & 0.928  (2.633) & 0.927  (2.85)\\
 & Dagum & 0.50 & 0.946  (2.983) & 0.966  (3.202) & 0.933  (2.715) & 0.955  (3.051)\\
 &  & 0.75 & 0.944  (3.468) & 0.938  (3.920) & 0.919  (3.658) & 0.941  (3.912)\\
 &  & 0.90 & 0.970  (5.844) & 0.946  (6.271) & 0.910  (6.430) & 0.945  (6.201)\\
\cline{2-7}
100 &  & 0.10 & 0.861  (6.315) & 0.901  (6.090) & 0.974  (5.100) & 0.883  (6.092)\\
 &  & 0.25 & 0.978  (6.024) & 0.979  (6.500) & 0.929  (5.778) & 0.954  (6.328)\\
 & Singh-Maddala & 0.50 & 0.949  (7.399) & 0.965  (7.561) & 0.939  (7.147) & 0.965  (7.337)\\
 &  & 0.75 & 0.918  (10.35) & 0.962  (10.26) & 0.922  (10.29 ) & 0.969  (10.11)\\
 &  & 0.90 & 0.859  (15.89) & 0.947  (15.65) & 0.847  (81.59) & 0.932  (16.96)\\
\cline{2-7}
 &  & 0.10 & 0.921  (0.714) & 0.952  (0.703) & 0.873  (0.706) & 0.924  (0.726)\\
 &  & 0.25 & 0.920  (0.558) & 0.961  (0.559) & 0.903  (0.526) & 0.961  (0.549)\\
 & Normal & 0.50 & 0.943  (0.496) & 0.958  (0.514) & 0.945  (0.481) & 0.971  (0.505)\\
 &  & 0.75 & 0.918  (0.546) & 0.965  (0.555) & 0.910  (0.535) & 0.963  (0.549)\\
 &  & 0.90 & 0.889  (0.704) & 0.945  (0.706) & 0.880  (0.789) & 0.912  (0.704)\\
\hline
\end{tabular}
\hspace*{-1cm}
 \end{table}

\begin{table}[H]
  \centering
  
    \caption{Empirical coverage probabilities and average widths of interval estimates of $x_p$ from binned data at nominal level 95\% for large sample sizes,each based on 1000 replications.}
\label{table:tab2}
\hspace*{-1cm}
\begin{tabular}{ccccccc}
\hline
\textit{n} & \textit{F} & \textit{p} & Histogram & Frequency Polygon & Linear Interpolation & GLD\\
\hline
 &  & 0.10 & 0.769  (0.091) & 0.805  (0.091) & 0.970  (0.078) & 0.853  (0.095)\\
 &  & 0.25 & 0.974  (0.070) & 0.972  (0.081) & 0.953  (0.073) & 0.952  (0.079)\\
 & Lognormal & 0.50 & 0.943  (0.082) & 0.969  (0.085) & 0.952  (0.075) & 0.944  (0.079)\\
 &  & 0.75 & 0.975  (0.094) & 0.954  (0.104) & 0.948  (0.101) & 0.971  (0.101)\\
 &  & 0.90 & 0.969  (0.130) & 0.954  (0.147) & 0.937  (0.145) & 0.976  (0.159)\\
\cline{2-7}

 &  & 0.10 & 0.619  (1.833) & 0.669  (1.930) & 0.967  (1.603) & 0.739  (1.956)\\
 &  & 0.25 & 0.899  (1.651) & 0.883  (1.896) & 0.942  (1.682) & 0.859  (1.816)\\
 & Dagum & 0.50 & 0.967  (1.906) & 0.976  (2.061) & 0.942  (1.730) & 0.966  (1.950)\\
 & & 0.75 & 0.856  (2.125) & 0.856  (2.495) & 0.954  (2.333) & 0.896  (2.490)\\
 & & 0.90 & 0.896  (3.263) & 0.841  (3.942) & 0.910  (3.950) & 0.924 ( 3.789)\\
\cline{2-7}

250 &  & 0.10 & 0.761  (4.026) & 0.761  (3.933) & 0.968  (3.253) & 0.804  (3.968)\\
&  & 0.25 & 0.963  (3.736) & 0.964  (4.117) & 0.955  (3.709) & 0.933  (4.027)\\
 & Singh-Maddala & 0.50 & 0.943  (4.922) & 0.960  (4.836) & 0.949  (4.490) & 0.953  (4.662)\\
 &  & 0.75 & 0.882  (7.024) & 0.963  (6.452) & 0.923  (6.342) & 0.951  (6.371)\\
 &  & 0.90 & 0.920  (10.09) & 0.949  (9.897) & 0.925  (9.612) & 0.930  (10.60)\\
\cline{2-7}

 && 0.10 & 0.947  (0.445) & 0.955  (0.451) & 0.935  (0.427) & 0.944  (0.447)\\
 &  & 0.25 & 0.925  (0.356) & 0.946  (0.353) & 0.934  (0.343) & 0.943  (0.350)\\
 & Normal & 0.50 & 0.942  (0.317) & 0.960  (0.326) & 0.944  (0.304) & 0.956  (0.320)\\
 &  & 0.75 & 0.950  (0.357) & 0.967  (0.352) & 0.935  (0.341) & 0.953  (0.350)\\
&  & 0.90 & 0.927  (0.435) & 0.941  (0.444) & 0.920 (0.425) & 0.937  (0.444)\\
\cline{1-7}

 &  & 0.10 & 0.554  (0.063) & 0.569  (0.064) & 0.963  (0.055) & 0.649  (0.067)\\
 &  & 0.25 & 0.959  (0.049) & 0.952  (0.057) & 0.965  (0.051) & 0.928  (0.056)\\
 & Lognormal & 0.50 & 0.940 (0.059) & 0.972  (0.060) & 0.948  (0.052) & 0.946  (0.056)\\
 &  & 0.75 & 0.969  (0.063) & 0.923  (0.073) & 0.941  (0.070) & 0.964  (0.071)\\
 &  & 0.90 & 0.981  (0.088) & 0.931  (0.104) & 0.933  (0.104) & 0.978  (0.113)\\
\cline{2-7}

 &  & 0.10 & 0.390  (1.274) & 0.390  (1.380) & 0.959  (1.136) & 0.529  (1.367)\\
 &  & 0.25 & 0.810  (1.173) & 0.790  (1.354) & 0.949  (1.189) & 0.793  (1.291)\\
 &  & 0.50 & 0.960  (1.364) & 0.965  (1.489) & 0.957  (1.229) & 0.961  (1.411)\\
 & Dagum & 0.75 & 0.760  (1.501) & 0.753  (1.764) & 0.966  (1.639) & 0.829  (1.764)\\
 &  & 0.90 & 0.765  (2.102) & 0.751  (2.768) & 0.945  (2.690) & 0.841  (2.626)\\
\cline{2-7}

500 &  & 0.10 & 0.520  (2.809) & 0.518  (2.769) & 0.973  (2.273) & 0.537  (2.822)\\
 &  & 0.25 & 0.964  (2.611) & 0.962  (2.906) & 0.96  (2.585) & 0.911  (2.844)\\
 & Singh-Maddala & 0.50 & 0.936  (3.551) & 0.954  (3.406) & 0.943  (3.154) & 0.940  (3.272)\\
 &  & 0.75 & 0.891  (5.158) & 0.929  (4.572) & 0.942  (4.502) & 0.932  (4.524)\\
 &  & 0.90 & 0.932  (7.255) & 0.945  (7.104) & 0.949  (6.848) & 0.960  (7.596)\\
\cline{2-7}

 &  & 0.10 & 0.921  (0.295) & 0.943  (0.313) & 0.938  (0.297) & 0.954  (0.311)\\
 &  & 0.25 & 0.952  (0.254) & 0.955  (0.246) & 0.950  (0.239) & 0.952  (0.246)\\
 & Normal & 0.50 & 0.955  (0.224) & 0.969  (0.229) & 0.957  (0.215) & 0.965  (0.226)\\
 &  & 0.75 & 0.946  (0.255) & 0.946  (0.247) & 0.948  (0.241) & 0.951  (0.246)\\
 &  & 0.90 & 0.941  (0.297) & 0.954  (0.314) & 0.954  (0.298) & 0.965  (0.310)\\
\hline
\end{tabular}
\hspace*{-1cm}
 \end{table}

As can be seen in Table \ref{table:tab1}, even for small sample sizes of $n=50$ and $n=100$, the GLD and frequency polygon methods work well, with coverage probabilities close to the nominal 0.95.  When the sample sizes increase as in Table \ref{table:tab2}, all the methods return reliable coverages except for when $p=0.1$ using the GLD, histogram and frequency polygon methods. For skewed distributions such as the Dagum, the linear interpolation method provides excellent results throughout, although the method require the additional information available in the form of bin means. For the other methods, coverage probabilities are poorer in the extremes for skewed distributions, which is not surprising given the difficultly faced in estimating the density at these points, especially based on limited information. When the distributions become more symmetric, all the methods produce reliable coverage results for sufficiently large sample sizes. Width of the intervals increase in the tails since the density estimates are smaller. All of the methods provide confidence intervals that are of similar average width.    

\begin{table}[H]
  \centering
  
    \caption{Empirical coverage probabilities and average widths of interval estimates of $x_p$ from binned data at nominal level 95\% for normal mixture distribution based on 1000 replications.}
\label{table:tab3}
\begin{tabular}{ccccc}
\hline
\textit{n} & \textit{p} & Histogram  & Frequency Polygon & Linear Interpolation\\
\hline
 & 0.10 & 0.951  (2.568) & 0.962  (2.466) & 0.921  (2.390)\\
 & 0.25 & 0.948  (2.287) & 0.968  (2.333) & 0.927  (2.040)\\
50 & 0.50 & 0.905  (3.876) & 0.885  (3.959) & 0.859  (5.989)\\
 & 0.75 & 0.966  (2.350) & 0.983  (2.418) & 0.879  (1.703)\\
 & 0.90 & 0.852  (1.790) & 0.939  (1.855) & 0.910  (1.320)\\
\cline{1-5}
 & 0.10 & 0.962  (1.700 ) & 0.959  (1.722) & 0.930  (1.555)\\
 & 0.25 & 0.939  (1.652) & 0.963  (1.643) & 0.917  (1.447)\\
100 & 0.50 & 0.946  (2.831) & 0.930  (2.842) & 0.901  (2.958)\\
 & 0.75 & 0.948  (1.659) & 0.966  (1.668) & 0.912  (1.214)\\
 & 0.90 & 0.774  (1.263) & 0.866  (1.294) & 0.941  (0.964)\\
\cline{1-5}
 & 0.10 & 0.966  (1.085) & 0.929  (1.104) & 0.947  (0.980)\\
 & 0.25 & 0.929  (1.042) & 0.953  (1.035) & 0.932  (0.921)\\
250 & 0.50 & 0.958  (1.760) & 0.964  (1.817) & 0.948  (1.775)\\
 & 0.75 & 0.915  (1.037) & 0.929  (1.043) & 0.961  (0.726)\\
 & 0.90 & 0.582  (0.781) & 0.617  (0.804) & 0.967  (0.614)\\
\cline{1-5}
 & 0.10 & 0.968  (0.722) & 0.883  (0.769) & 0.950  (0.684)\\
 & 0.25 & 0.951  (0.719) & 0.951  (0.715) & 0.945  (0.646)\\
500 & 0.50 & 0.959  (1.219) & 0.960  (1.260) & 0.948  (1.216)\\
 & 0.75 & 0.888  (0.696) & 0.910  (0.699) & 0.953  (0.508)\\
 & 0.90 & 0.449  (0.503) & 0.453  (0.533) & 0.972  (0.431)\\
\hline
\end{tabular}

 \end{table}

In Table \ref{table:tab3} we provide coverage results for the normal mixture distribution with parameters $\mu= 10,4$, $\sigma=1,2$ and weights $w=0.4,0.6$. Quantiles which are lying near the connecting point of the two distributions shows some poorer coverages due to sensitivity of the density estimates near this location.  Again converages are typically close to nominal in the non-extremes.  In particular, coverage is excellent for the median for moderate to large sample sizes.  

\begin{figure}[h]
\centering
\includegraphics[height=11cm,width=\textwidth]{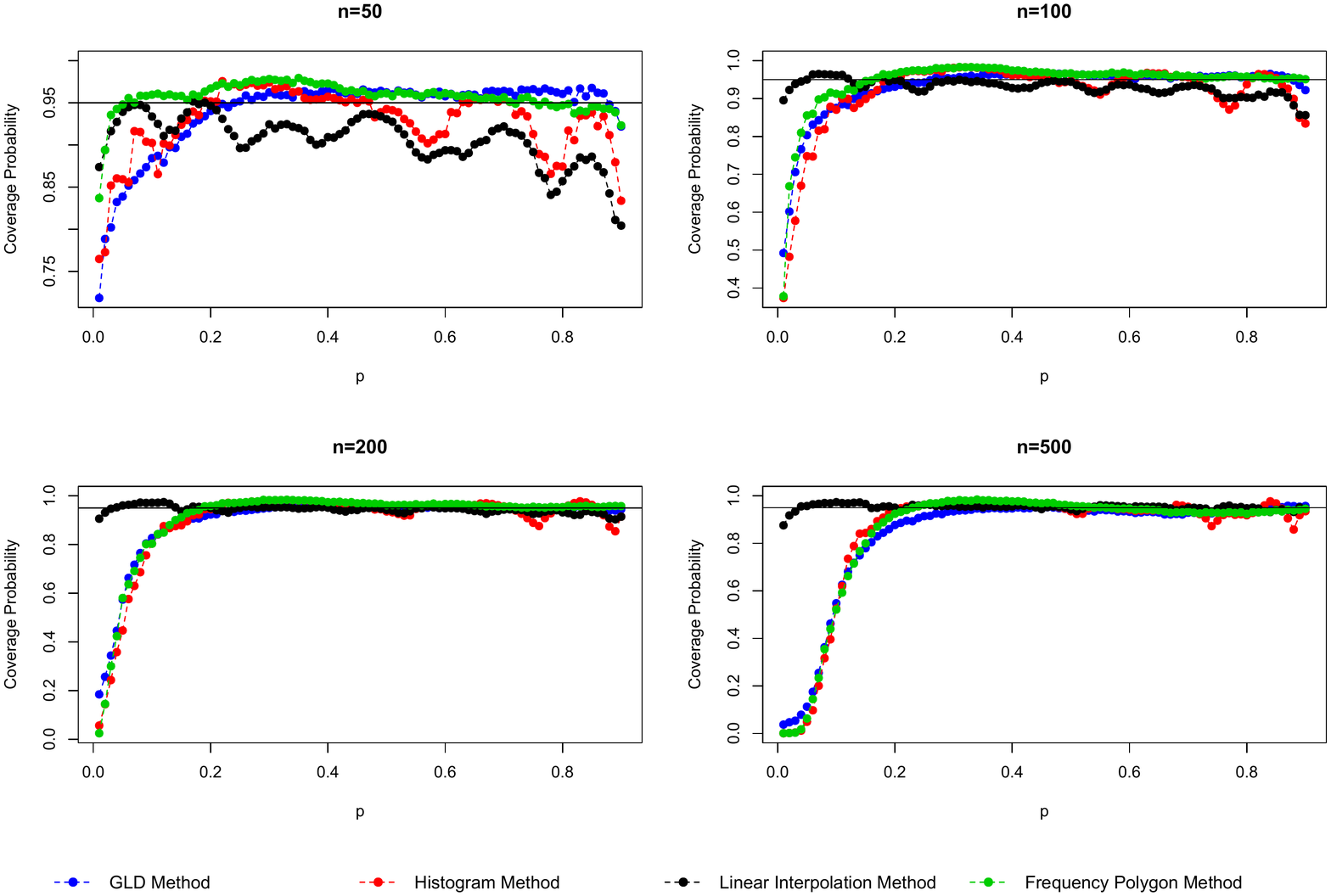}
\caption{Simulated coverage probabilities for the interval estimation of $x_p$ from binned data generated from the Singh-Maddala distribution with parameters from fitted 1980 US family income.}
\label{fig:fig4}
\end{figure}

In Figure \ref{fig:fig4}, a more detailed simulation is presented comparing the methods for each sample size over a grid of size 100 for $p$ in $(0,\ 1)$. The coverage curves depict that the GLD, histogram and frequency polygon methods are not reliable in estimating intervals in the lower extremes. On the other hand, if the bin means are available, the linear interpolation method is more reliable in estimating even these smaller quantiles for sufficiently large sample sizes. However, very reliable coverages can be achieved for not too small quantiles, despite the limited information available.   

\subsection{Limitations} \label{Limitations}

As described in \cite{cowell2011measuring}, the density estimate  by linear interpolation method is non negative only if the interval mean falls in middle third of the interval.  Additionally, poor density estimates may result using all of the methods when a large amount of the data falls in only a few bins. This may happen for decaying type distributions unless the number of intervals are sufficiently large.  As an example we provide coverage probabilities the median interval estimator based on data simulated from an exponential distribution in Table \ref{table:tab4}.   Coverage can be poor when the number of bins is small which, due to the nature of this distribution, can mean that most of the data can fall into only one or two bins.  However, coverage improves as the number of bins increases.

\begin{table}[H]
\centering
\caption{Empirical coverage probabilities and average widths of interval estimates of $x_{0.5}$ from binned data for different number of intervals at nominal level 95\% for Exponential Distribution (rate =1) based on 1000 replications. }
\label{table:tab4}
\hspace*{-0.5cm}
\begin{tabular}{cccccc}
\hline
\multicolumn{1}{c}{ } & \multicolumn{1}{c}{ } & \multicolumn{3}{c}{Methods} \\
\cline{3-6}
\textit{n} & \# of intervals & Histogram & Frequency Polygon & Linear Interpolation & GLD\\
\hline
 & 5 & 0.892 (0.484) & 0.939 (0.562) & 0.917 (0.550) & 0.939 (0.547)\\
 & 10 & 0.892 (0.572) & 0.959 (0.560) & 0.893 (0.587) & 0.920 (0.531)\\
50 & 15 & 0.882 (0.575) & 0.933 (0.569) & 0.872 (0.672) & 0.927 (0.534)\\
 & 20 & 0.910 (0.559) & 0.933 (0.562) & 0.860 (0.806) & 0.940 (0.532)\\
\cline{1-6}
 & 5 & 0.804 (0.331) & 0.805 (0.401) & 0.945 (0.384) & 0.876 (0.395)\\
 & 10 & 0.950 (0.403) & 0.960 (0.393) & 0.949 (0.384) & 0.954 (0.382)\\
100 & 15 & 0.945 (0.413) & 0.963 (0.398) & 0.941 (0.390) & 0.951 (0.382)\\
 & 20 & 0.919 (0.397) & 0.944 (0.395) & 0.893 (0.392) & 0.935 (0.381)\\
\cline{1-6}
 & 5 & 0.458 (0.215) & 0.373 (0.259) & 0.947 (0.237) & 0.653 (0.252)\\
 & 10 & 0.857 (0.249) & 0.827 (0.250) & 0.964 (0.246) & 0.916 (0.247)\\
250 & 15 & 0.939 (0.262) & 0.948 (0.250) & 0.941 (0.247) & 0.938 (0.245)\\
 & 20 & 0.937 (0.258) & 0.956 (0.25) & 0.940 (0.247) & 0.949 (0.244)\\
\cline{1-6}
 & 5 & 0.077 (0.162) & 0.036 (0.190) & 0.949 (0.166) & 0.233 (0.179)\\
 & 10 & 0.631 (0.168) & 0.609 (0.176) & 0.955 (0.172) & 0.717 (0.175)\\
500 & 15 & 0.914 (0.184) & 0.919 (0.178) & 0.966 (0.175) & 0.910 (0.174)\\
 & 20 & 0.928 (0.183) & 0.935 (0.177) & 0.953 (0.174) & 0.922 (0.174)\\
\hline
\end{tabular}
\hspace*{-0.5cm}
\end{table}

\subsection{Example 1: Glucose effectiveness for young and healthy Caucasian men and women } \label{Example 1}

We now return to the motivating example presented in Section \ref{Motivating Example} comparing Glucose effectiveness between males and females.  

\begin{table}[H]
  \centering
  
    \caption{Confidence Intervals for difference of quartiles between males and females for the Glucose effectiveness example.}
\label{table:tab5}
\hspace*{-0.5cm}
\begin{tabular}{lllll}
\cline{2-4}
\multicolumn{1}{c}{} & \multicolumn{1}{c}{GLD Method} & \multicolumn{1}{c}{Histogram Method} & \multicolumn{1}{c}{Frequency Polygon Method} &  \\ 
\cline{1-4}
\multicolumn{1}{c}{CI for difference of $x_{0.25}$ } & \multicolumn{1}{c}{(0.122, 0.463)}  & \multicolumn{1}{c}{(0.128, 0.482)} & \multicolumn{1}{c}{(0.137, 0.496)}\\ 
\multicolumn{1}{c}{CI for difference of $x_{0.5}$  } & \multicolumn{1}{c}{(0.145, 0.450)} & \multicolumn{1}{c}{(0.141, 0.453)}  & \multicolumn{1}{c}{(0.144, 0.455)}\\ 
\multicolumn{1}{c}{CI for difference of $x_{0.75}$  }& \multicolumn{1}{c}{(0.131, 0.455)} & \multicolumn{1}{c}{(0.149, 0.462)}  & \multicolumn{1}{c}{(0.128, 0.461)}\\ 
\cline{1-4}
\end{tabular}
 \end{table}

Table \ref{table:tab5} contains the confidence interval estimates for the difference of quartiles comparing the two groups using \eqref{eq3} and only the information available in the histograms in Figure \ref{fig1}.  As can be seen, the confidence intervals generated from each method are similar to one another and reasonably precise. Further, all confidence intervals clearly suggest that there is a significant difference between the two groups for glucose effectiveness. These results align with the initial findings of the \cite{clausen1996insulin} using the full data set that there is a difference between the two groups for glucose effectiveness.            

\subsection{Example 2: Solubility distribution for quantified proteins } \label{Example 2}

\begin{figure}[H]
\centering
\includegraphics[height=8cm,width=12cm]{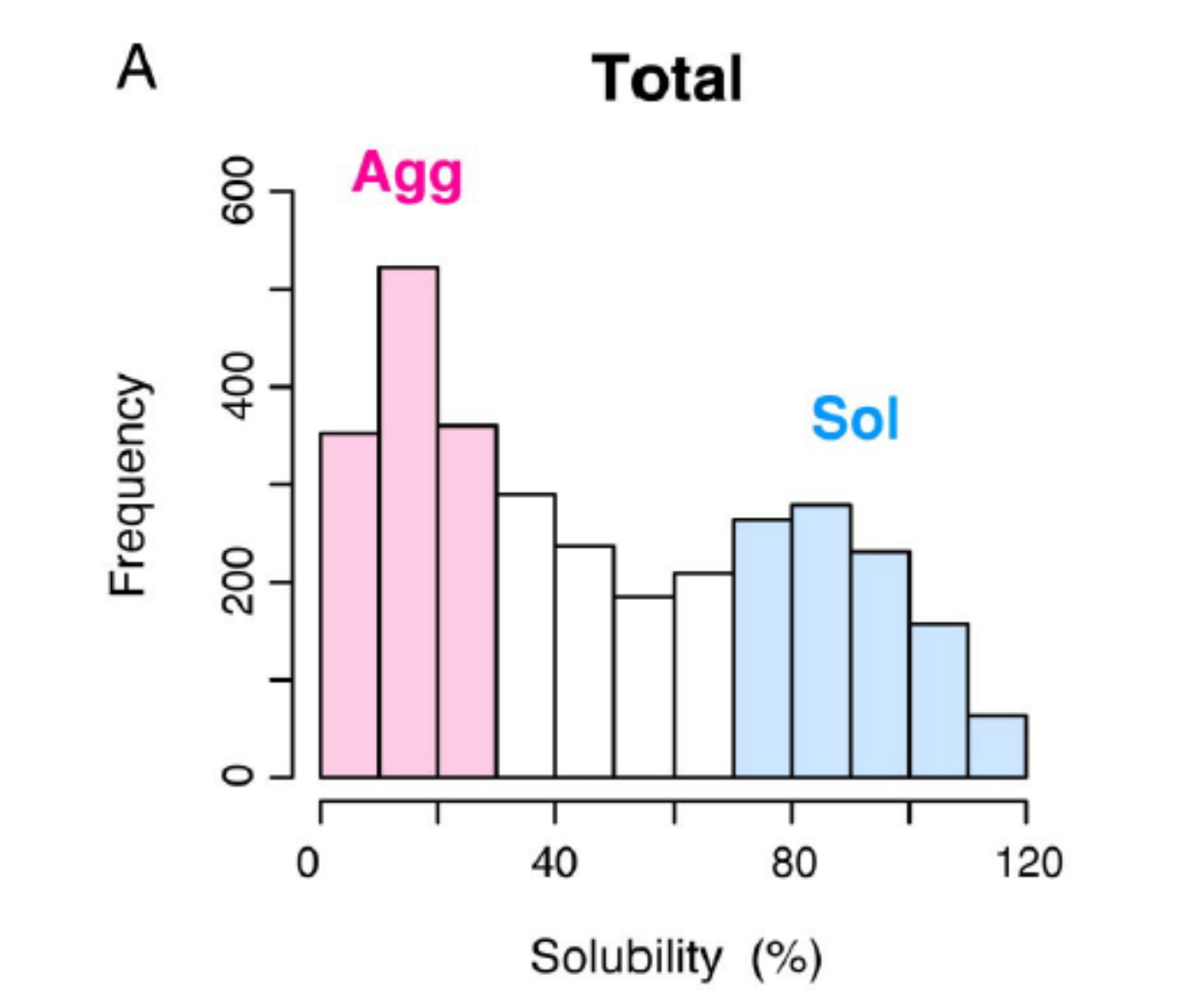}
\caption{ histogram of
solubility for the 3,173 quantified proteins. The proteins with solubilities 30\% and 70\% were defined as the aggregation-prone (Agg, colored pink) and soluble (Sol, colored blue) groups, respectively. Reprinted from "Bimodal protein solubility distribution revealed by an aggregation analysis of the entire ensemble of Escherichia coli proteins," by T. Niwaa, B.W. Yinga, K. Saitoa, W. Jinc, S. Takadac, T. Uedaa, and H. Taguchia, 2009, Proceedings of the National Academy of Sciences, 106, p. 4201-4206. Reprinted with permission.}
\label{fig5}
\end{figure}

In this example we present a bimodal histogram in Figure \ref{fig5} published by \cite{niwa2009bimodal} for solubility levels from 3,173 translated proteins. The histogram highlights two groups; namely the aggregation prone group (Agg, defined as $<30\%$) and the highly soluble group (Sol, defined as $>70\%$) depending on the properties they hold. 

\begin{table}[H]
  \centering
  
    \caption{Confidence Intervals for quartiles from the bimodal solubilities example.}
\label{table:tab6}
\begin{tabular}{ccc}
\cline{2-3}
 & Histogram Method & Frequency Polygon Method\\
\hline
CI for $x_{0.25}$ & (17.433 , 19.243) & (17.512 , 19.536)\\
CI for $x_{0.5}$ & (39.560 , 44.103) & (39.783 , 44.066)\\
CI for $x_{0.75}$ & (75.982 , 79.674) & (75.990 , 79.618)\\
\hline
\end{tabular}

 \end{table}

If intervals of quantiles are of interest then these can be obtained using the methods presented earlier.  We provide the quartile interval estimates in Table \ref{table:tab6} generated using the histogram and the frequency polygon methods and the limited information available in Figure \ref{fig5}. As can be seen the confidence intervals generated from both methods are almost identical.

\section{A Shiny application for calculation of the intervals} \label{A Shiny application for calculation of the intervals}
We have also created a Shiny \cite{shiny} application that can be accessed at
\begin{center}
\url{https://lukeprendergast.shinyapps.io/histCIs/}
\end{center}
The easiest way to use the application is to upload a .csv file that contains at least the lower and upper bounds for the bins and the frequencies.  However, the table in the application is interactive and rows can be added and deleted.  Once the data is uploaded the user can press the `Save Table' button and a set of options will appear (i.e. method, confidence level etc.).  Once the desired options are selected the `Calculate Now' button can be pressed and the interval will be calculated.  Improvements to this application will continue to be made.

\section{Discussion} \label{Discussion}

Our results have shown that reliable confidence intervals can be obtained for a wide range of quantiles when only grouped data (i.e. such as histograms) are available.  While it is not advisable to attempt to obtain a confidence interval for a very small quantile, if bin means are available along with the frequencies then the linear interpolation approach based on density estimator of \cite{lyon2016advantages} can provide reliable intervals if the sample size is not small. For unimodal distributions, the median is likely to be the quantile that is of most interest, and all of the methods provide excellent results for most distributions that we have studied.  A word of caution may be necessary if most of the sample falls into a very small number of bins.  For example, for decaying distributions such as the exponential, there may be situations where the majority of data falls within just the first few bins rendering a density estimate near the median unreliable.  However, for these types of distributions intervals for larger quantiles provide better coverage.  The results presented here can be extended to other functions of quantiles where the variance depends on the sample size, density function and quantiles.  For example, \cite{Prendergast2017} considered confidence intervals estimators for the ratio of dependent quantiles.  Approximate confidence intervals from histograms for these ratios could also be found since the covariance between the quantiles depends only on the aforementioned quantities.   In summary, when data is summarized in a grouped format we have provided ways in which confidence intervals can be obtained for quantiles.  This means that researchers can interrogate published research that may only, for example, have a histogram available.  

\bibliographystyle{authordate4}
\bibliography{ms.bib}

\end{document}